\documentclass[showpacs,twocolumn,preprintnumbers,amsmath,amssymb,pra]{revtex4}
\usepackage{graphicx}
\usepackage{subfigure}

\begin{document}


\preprint{ND Atomic Physics/04/06}

\title{Third-order many-body perturbation theory calculations\\
for the beryllium and magnesium isoelectronic sequences}

\author{H.\,C. Ho}\email{hho1@nd.edu}\thanks{present address: NCTS (Physics Division), 101 Sec.\ 2 Kuang Fu Road, Hsinchu, Taiwan, R.O.C.}
\author{W.\,R. Johnson}\email{johnson@nd.edu}
\affiliation{Department of Physics, University of Notre Dame,
Notre Dame, Indiana 46556, USA}

\author{S.\,A. Blundell}\email{steven.blundell@cea.fr}
\affiliation{D\'epartement de Recherche Fondamentale sur la
Mati\`ere Condens\'ee, CEA--Grenoble/DSM\\
17 rue des Martyrs, F--38054 Grenoble Cedex 9, France}

\author{M.\,S. Safronova}\email{msafrono@udel.edu}
\affiliation{Department of Physics, University of Delaware,
Newark, Delaware 19716, USA}

\date{\today}

\begin{abstract}

Relativistic third-order MBPT is applied to obtain energies of ions with
two valence electrons in the no virtual-pair approximation (NVPA). A
total of 302 third-order Goldstone diagrams are organized into 12
one-body and 23 two-body terms. Only third-order two-body terms and
diagrams are presented here, owing to the fact that the one-body terms
are identical to the previously studied third-order terms in monovalent
ions. Dominant classes of diagrams are identified. The model potential
is a Dirac-Hartree-Fock $V^{N-2}$ potential, and B-spline basis
functions in a cavity of finite radius are employed in the numerical
calculations. The Breit interaction is taken into account through second
order of perturbation theory and the lowest-order Lamb shift is also
evaluated. Sample calculations are performed for berylliumlike ions with
$Z$ = 4--7, and for the magnesiumlike ion P\,IV.  The third-order
energies are in excellent agreement with measurement with an accuracy at
0.2\% level for the cases considered. Comparisons are made with previous
second-order MBPT results and with other calculations. The third-order
energy correction is shown to be significant, improving second-order
correlation energies by an order of magnitude.

\end{abstract}



\maketitle

\section{Introduction}

The development of relativistic MBPT in recent decades has been motivated in part
by the need for accurate theoretical amplitudes of parity non-conserving (PNC)
transitions in heavy monovalent atoms such as cesium and francium.
Applications of the theoretical methods developed to treat atomic PNC include
support of atomic clock development, tests of QED in precision spectroscopy
of highly-stripped ions, searches for time variation in the fine-structure constant,
and providing precise astrophysical data.

Although nonrelativistic studies~\cite{key-Ivanova1967,key-Safronova1969,%
key-Ivanova1975,key-Chang1989,key-Chang1986,key-Fischer1997,key-Galvez2003}
and relativistic HF calculations~\cite{key-Cheng1979,key-Ynnerman1995,key-Jonsson1997}
for divalent atoms and ions have been done for many years,
only recently have relativistic many-body
calculations been reported. As examples, we note that all-order relativistic MBPT calculations
for transitions in berylliumlike ions with $Z=26$ and $42$ were carried out by
\citet{key-Lindroth1992},
while large-scale configuration-interaction (CI) calculations for transitions in C\, III
were performed by \citet{key-Chen2001}.  Relativistic many-body calculations for magnesiumlike
ions include the CI calculations of states in the $n=3$ complex by
\citet{key-Chen1997}, and mixed CI-MBPT calculations of excitation energies in neutral Mg by \citet{key-Savukov2002}.

Second-order relativistic MBPT was applied
to Be-like ions by \citet{key-Safronova1996} and energies were found to
be accurate at the 2\% level.
In this paper, we extend relativistic MBPT for divalent atoms and ions to third order.
We give a detailed treatment of the two-body terms here; the one-body terms are identical to those for monovalent systems and are
discussed in detail by \citet{key-Blundell1990}.
The long-range goal of the present research is to extend the relativistic singles-doubles coupled-cluster (SDCC)
formalism to atoms and ions with two valence electrons. The present calculations permit us to identify and
evaluate those third-order terms missing from the SDCC expansion.

\section{Theoretical Method}

The model potential for our MBPT calculation is the Dirac-Hartree-Fock
$V^{N-2}$ potential. The Dirac-Coulomb-Breit Hamiltonian for an $N$-electron
atom is $H=H_{0}+{V}$, where\[
H_{0}=\sum_{i=1}^{N}\left[h_{\textrm{D}}(i)+u\left(r_{i}\right)\right],\]
 and the Dirac Hamiltonian is\[
h_{\textrm{D}}=c\mbox{\boldmath$\alpha\cdot$}\mathbf{p}+\beta c^{2}+ V_\mathrm{nuc}(r),\]
where $ V_\mathrm{nuc}$ is obtained assuming a Fermi nuclear charge distribution.
 All equations are in atomic units. The perturbation is\[
{V}=\sum_{i<j=1}^{N}\left(\frac{1}{r_{ij}}+b_{ij}\right)-\sum_{i=1}^{N}u\left(r_{i}\right),\]
 where $b_{ij}$ is the Breit interaction\[
b_{ij}=-\frac{1}{2r_{ij}}\left[\mbox{\boldmath$\alpha_{i}$}\cdot\mbox{\boldmath$\alpha_{j}$}+\frac{\left(\mbox{\boldmath$\alpha_{i}$}\cdot\mbox{\boldmath$r_{ij}$}\right)\left(\mbox{\boldmath$\alpha_{j}$}\cdot\mbox{\boldmath$r_{ij}$}\right)}{r_{ij}^{2}}\right].\]
 In the no virtual-pair approximation, the excitations are limited
to positive-energy eigenstates of $H$.~\cite{key-Sucher1980,key-Mittleman1971,%
key-Mittleman1972,key-Mittleman1981}

The eigenstates of a divalent system having angular momentum $(J,M)$
are described by the coupled states
\begin{equation}
\Phi_{JM}(vw)  =  \eta_{vw}\sum_{m_{v}m_{w}}\left\langle j_{v}j_{w},m_{v}m_{w}|JM\right\rangle
                   a_{v}^{\dagger}a_{w}^{\dagger}\left|0\right\rangle,\label{eq:AM-states}
\end{equation}
where $\left|0\right\rangle $ represents the ground state of the
ionic core and $\eta_{vw}$ is the normalization constant
\[
\eta_{vw}=\left\{ \begin{array}{cl}
1 & \textrm{for }v\neq w\\
\frac{1}{\sqrt{2}} & \textrm{for }v=w\end{array}\right..
\]
 Here $v$ and $w$ specify the corresponding one-electron states
with quantum numbers $\left(n_{v},l_{v},j_{v},m_{v}\right)$ and $\left(n_{w},l_{w},j_{w},m_{w}\right)$.
The model space $P$ is defined by the set of total angular-momentum
states~(\ref{eq:AM-states}); the model-space projection operator is
\[
\mathbf{P}=\sum_{\stackrel{JM}
{{\scriptscriptstyle (v\leq w)}}}\left|\Phi_{JM}(vw)\right\rangle \left\langle \Phi_{JM}(vw)\right|.
\]
 The orthogonal-space operator $\mathbf{Q}$ is simply $\mathbf{1}\mbox{\boldmath$-$}\mathbf{P}$.

The wave operator $\Omega$ is found by solving the generalized Bloch
equation~\cite{key-Lindgren-Morrison1986}
\[
\left[\Omega,H_{0}\right]\mathbf{P}=({V}\Omega-\Omega\mathbf{P}{V}\Omega)\mathbf{P}.
\]
 The effective Hamiltonian is given in terms of the wave operator\[
H_{\textrm{eff}}=PH_{0}P+P{V}\Omega P.\]

\subsection{Effective Hamiltonian}

\begin{figure}

\centering
     \subfigure{\includegraphics[scale=0.4]{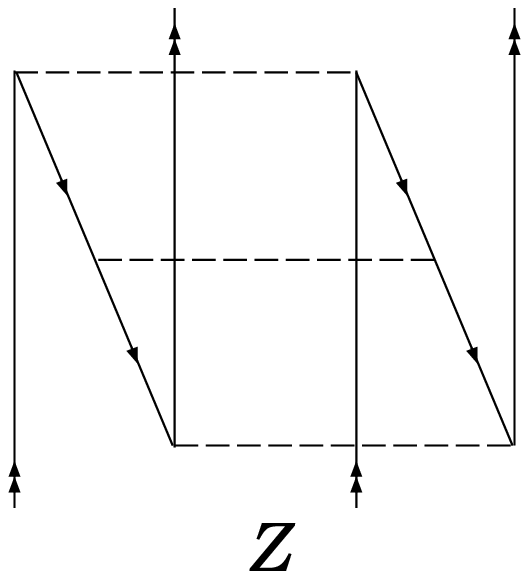}}\hspace{0.05in}
     \subfigure{\includegraphics[scale=0.4]{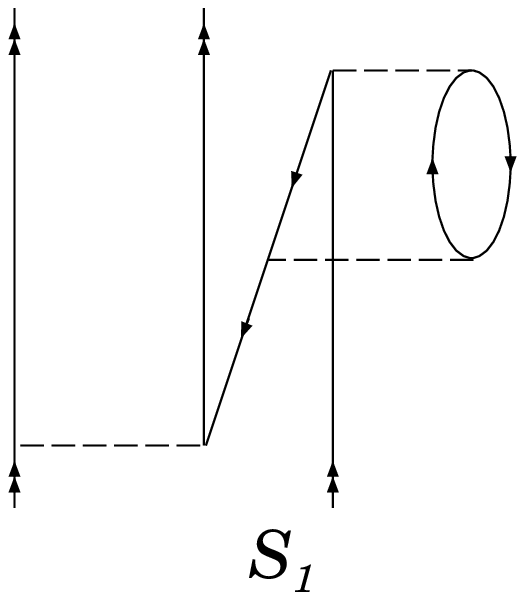}}\hspace{0.05in}
     \subfigure{\includegraphics[scale=0.4]{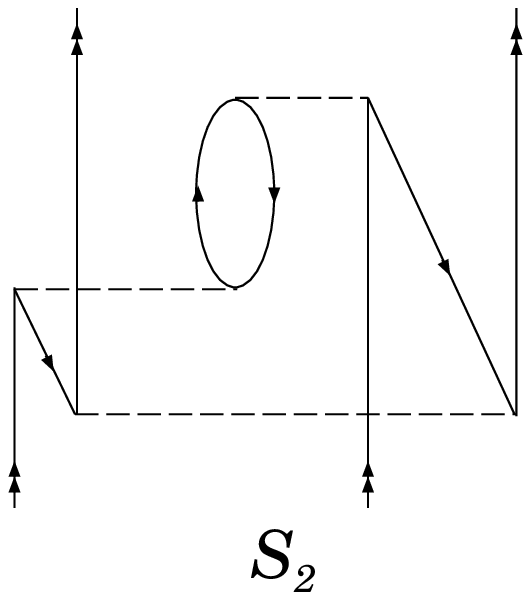}}\vspace{-0.12in}\\
     \subfigure{\includegraphics[scale=0.4]{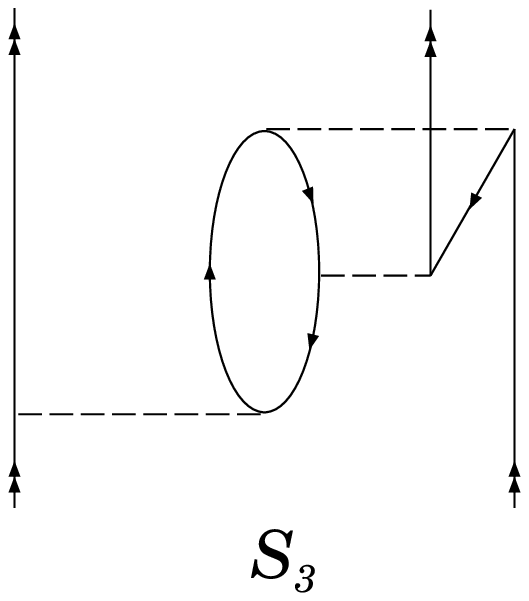}}\hspace{0.05in}
     \subfigure{\includegraphics[scale=0.4]{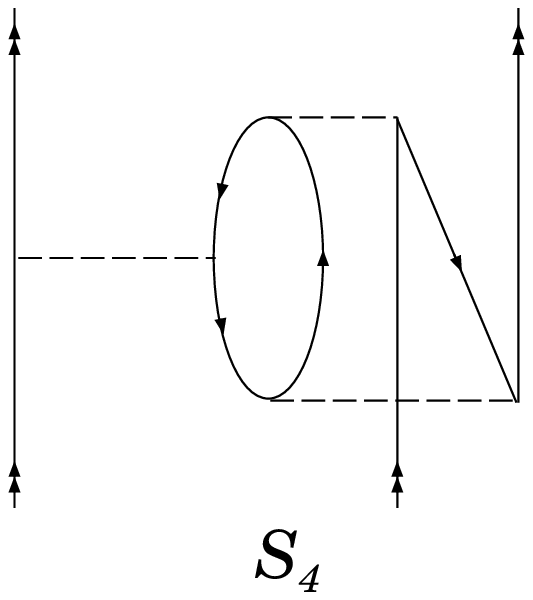}}\hspace{0.05in}
     \subfigure{\includegraphics[scale=0.4]{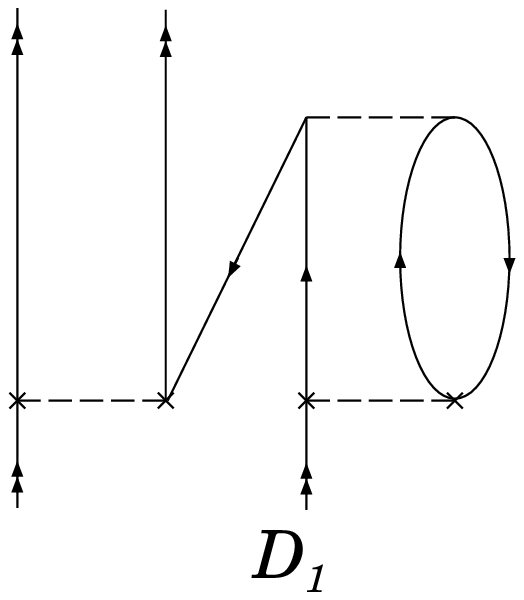}}\vspace{-0.12in}\\
     \subfigure{\includegraphics[scale=0.4]{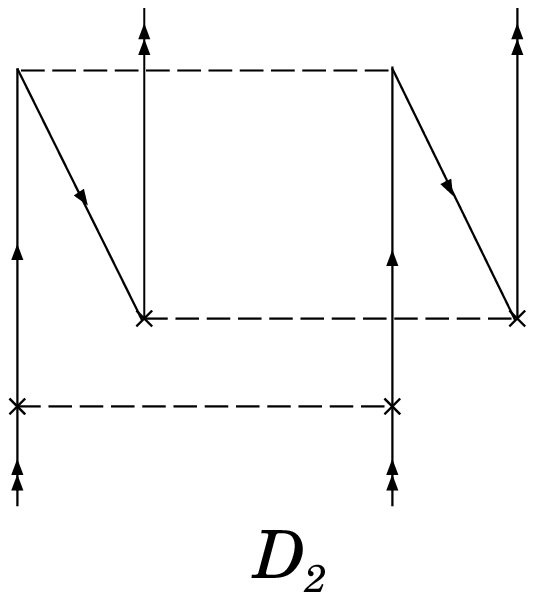}}\hspace{0.05in}
     \subfigure{\includegraphics[scale=0.4]{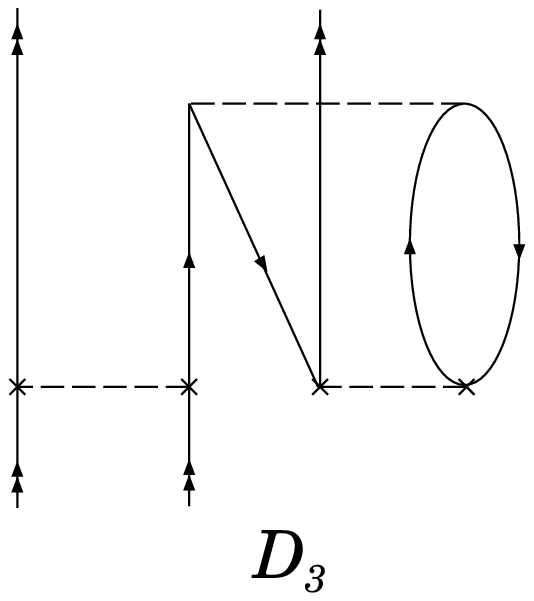}}\hspace{0.05in}
     \subfigure{\includegraphics[scale=0.4]{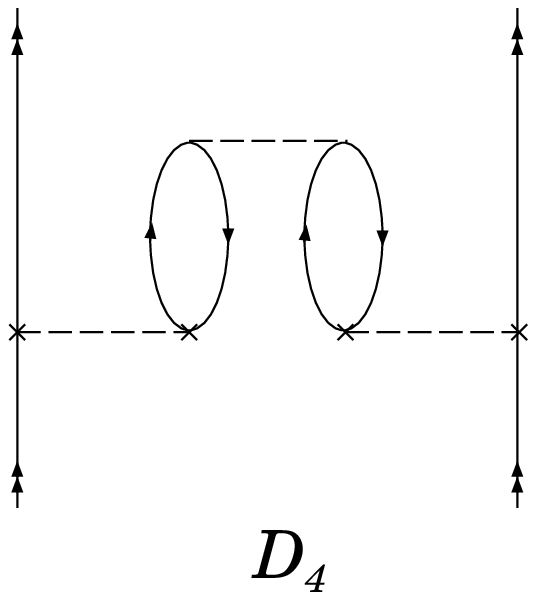}}\vspace{-0.12in}\\
    \subfigure{\includegraphics[scale=0.4]{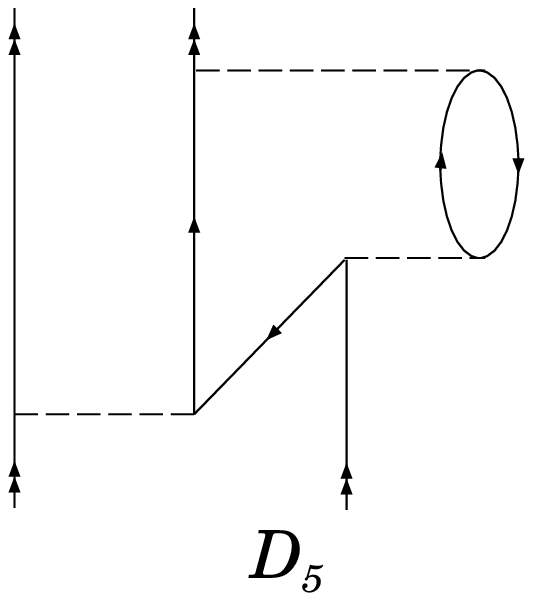}}\hspace{0.05in}
    \subfigure{\includegraphics[scale=0.4]{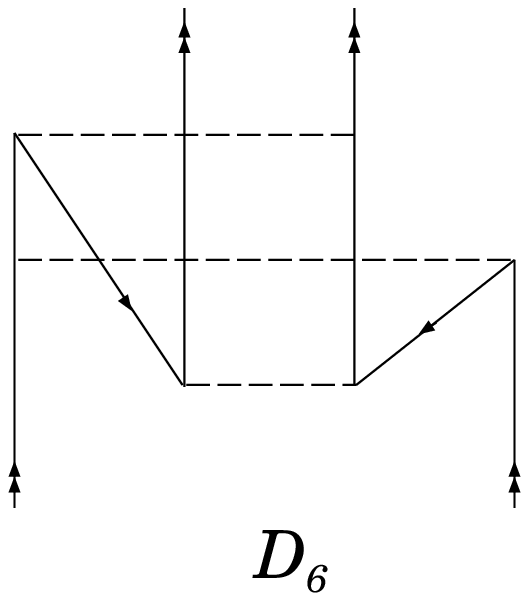}}\hspace{0.05in}
    \subfigure{\includegraphics[scale=0.4]{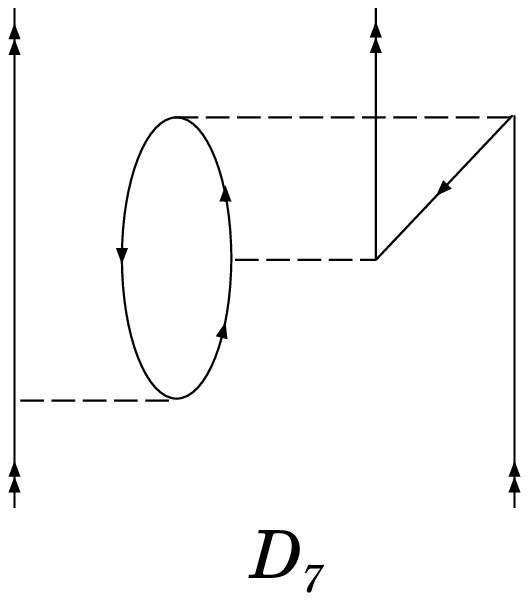}}\vspace{-0.12in}\\
    \subfigure{\includegraphics[scale=0.4]{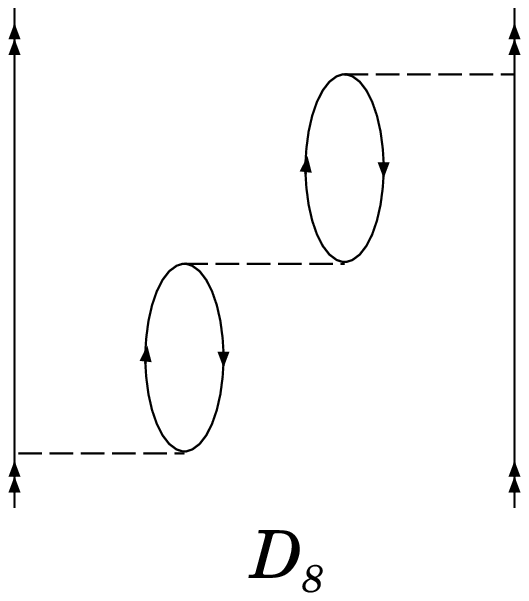}}\hspace{0.05in}
    \subfigure{\includegraphics[scale=0.4]{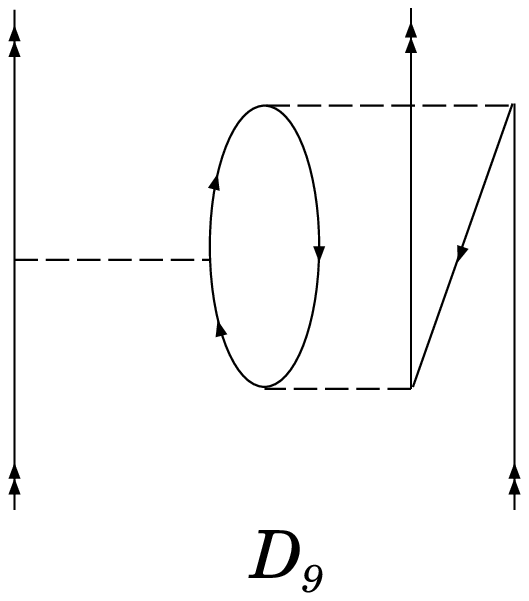}}\hspace{0.05in}
    \subfigure{\includegraphics[scale=0.4]{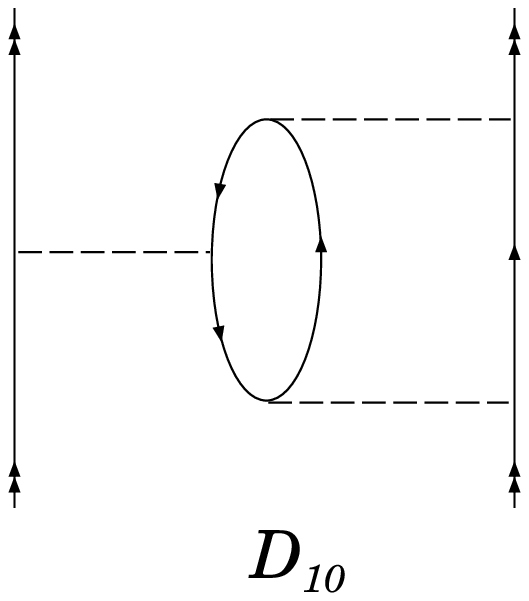}}\vspace{-0.12in}\\
    \subfigure{\includegraphics[scale=0.4]{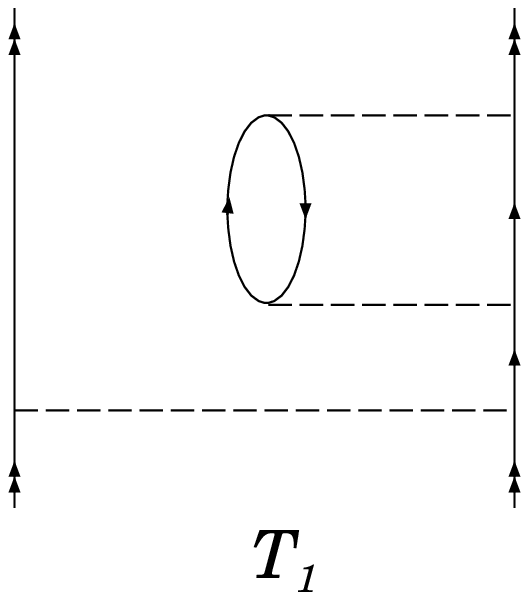}}\hspace{0.05in}
    \subfigure{\includegraphics[scale=0.4]{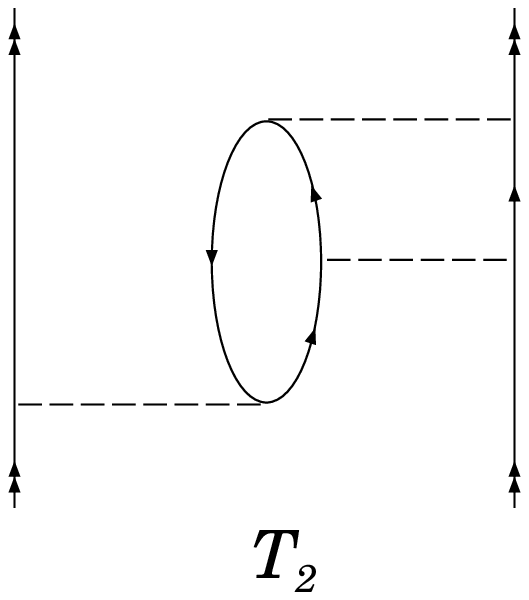}}\hspace{0.05in}
    \subfigure{\includegraphics[scale=0.4]{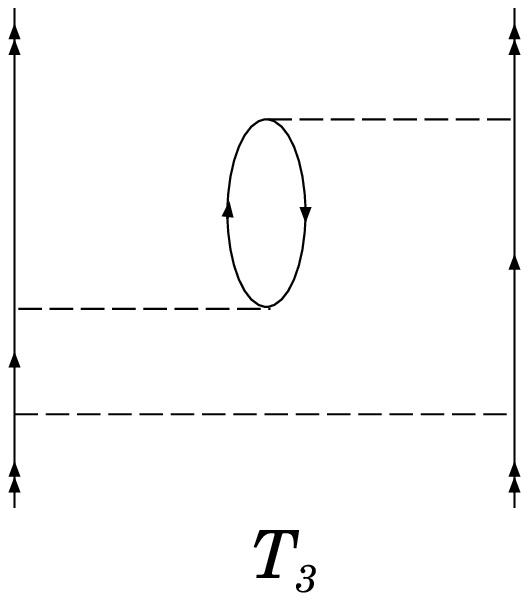}}\vspace{-0.12in}\\
    \subfigure{\includegraphics[scale=0.4]{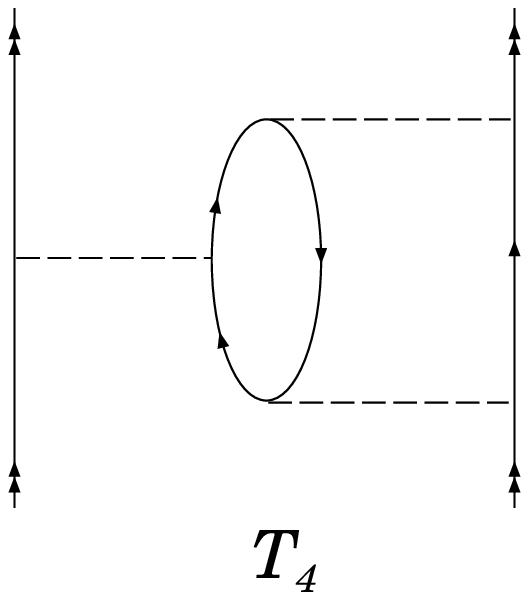}}\hspace{0.05in}
    \subfigure{\includegraphics[scale=0.4]{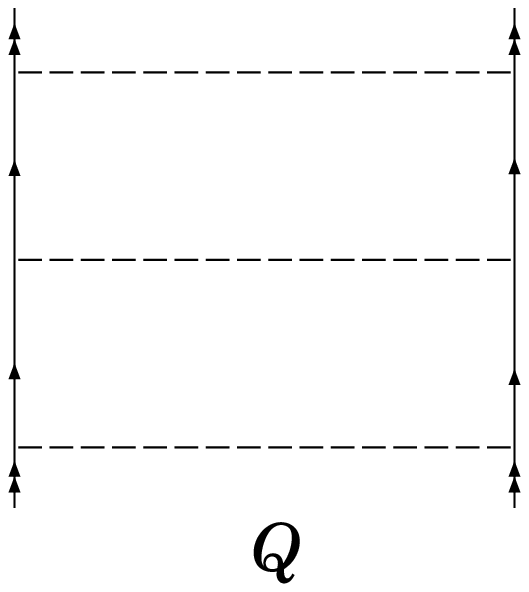}}\vspace{0.05in}
    \subfigure{\includegraphics[scale=0.4]{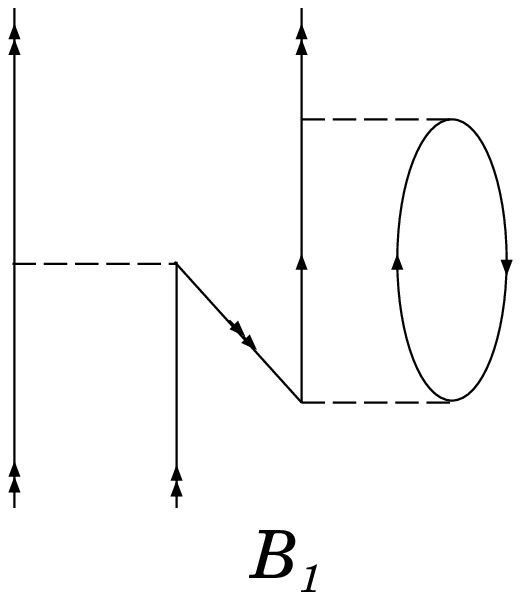}}\vspace{-0.12in}\\
    \subfigure{\includegraphics[scale=0.4]{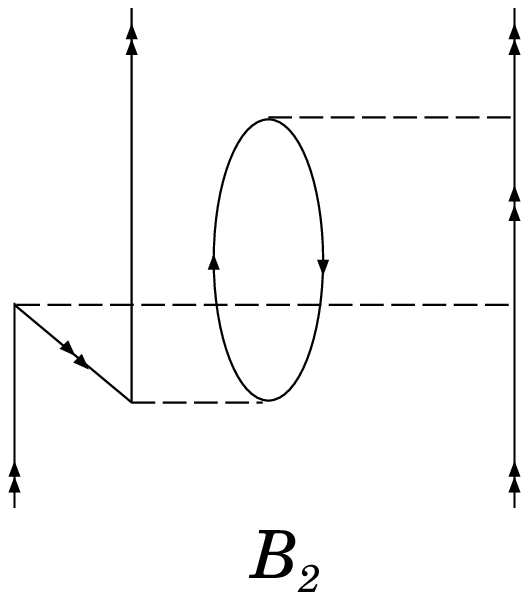}}\hspace{0.05in}
    \subfigure{\includegraphics[scale=0.4]{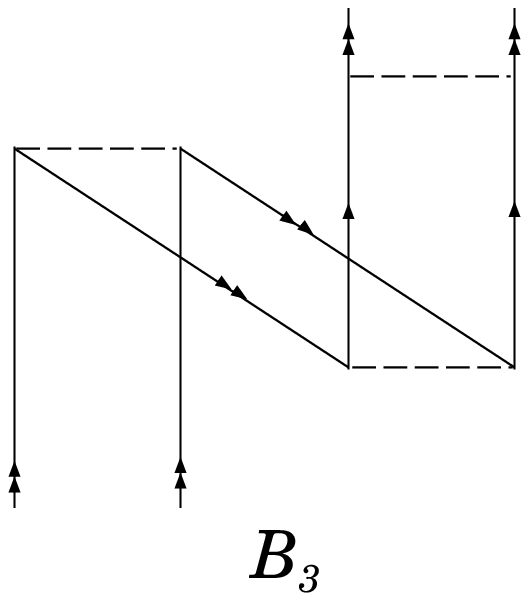}}\\
\caption{Third-order Goldstone diagrams (two-body part).}\label{cap:Third-II}
\end{figure}

We first find the configuration-weight vector by diagonalizing the
Hamiltonian $H=H_{\textrm{eff}}^{(0)}+H_{\textrm{eff}}^{(1)}$ using
total angular-momentum eigenstates~(\ref{eq:AM-states}) as basis.
Higher-order energies are obtained by operating the effective Hamiltonian
of the corresponding order on the configuration-weight vector. For
simplicity, matrix elements of the effective Hamiltonian are given
for the uncoupled states, $\left|vw\right\rangle
\equiv a_{{v}}^{\dagger}a_{{w}}^{\dagger}\left|0\right\rangle $
and $\left|v^{\prime}w^{\prime}\right\rangle $. In third order,
each element consists of 12 one-body and 23 two-body terms.
They represent a total of 84 one-body and 218 two-body Goldstone
diagrams. The multiplications of Clebsch-Gordan coefficients and the
summations over magnetic quantum numbers are carried out during angular
decomposition. Only the two-body part of the third order is discussed
here as the one-body part and complete second-order results are already
presented in Refs.~\cite{key-Blundell1990} and \cite{key-Safronova1996}.%
{\small\begin{eqnarray*}
Z     & = & \sum_{abcd}\frac{g_{cdwv}g_{abcd}\tilde{g}_{w'v'ab}}{\left(\varepsilon_{ab}-\varepsilon_{v'w'}\right)\left(\varepsilon_{cd}-\varepsilon_{v'w'}\right)}\\
S_{1} & = & \sum_{abcm}\frac{g_{acmw}\tilde{g}_{mbac}\tilde{g}_{v'w'vb}}{\left(\varepsilon_{bv}-\varepsilon_{v'w'}\right)\left(\varepsilon_{acv}-\varepsilon_{mv'w'}\right)}\\
&  & \times\left(1+\left[v\leftrightarrow w,\,v'\leftrightarrow w'\right]\right)+\text{c.c.}\\
S_{2} & = & \sum_{abcm}\frac{\tilde{g}_{abwm}\tilde{g}_{mcbv}\tilde{g}_{v'w'ca}}{\left(\varepsilon_{ac}-\varepsilon_{v'w'}\right)\left(\varepsilon_{abv}-\varepsilon_{mv'w'}\right)}\\
&  & \times\left(1+\left[v\leftrightarrow w,\,v'\leftrightarrow w'\right]\right)+\text{c.c.}\\
S_{3} & = & \sum_{abcm}\frac{g_{abwm}\tilde{g}_{w'cab}\tilde{g}_{v'mvc}}{\left(\varepsilon_{cv}-\varepsilon_{mv'}\right)\left(\varepsilon_{abv}-\varepsilon_{mv'w'}\right)}\\
&  & \times(1-[v\leftrightarrow w])\left(1-\left[v'\leftrightarrow w'\right]\right)+\text{c.c.}\\
S_{4} & = & \sum_{abcm}\frac{\tilde{g}_{abwm}\tilde{g}_{v'cvb}\tilde{g}_{w'mac}}{\left(\varepsilon_{ac}-\varepsilon_{mw'}\right)\left(\varepsilon_{abv}-\varepsilon_{mv'w'}\right)}\\
&  & \times(1-[v\leftrightarrow w])\left(1-\left[v'\leftrightarrow w'\right]\right)\\
D_{1} & = & -\sum_{abmn}\frac{g_{abmn}\tilde{g}_{mnwb}\tilde{g}_{v'w'va}}{\left(\varepsilon_{bw}-\varepsilon_{mn}\right)\left(\varepsilon_{av}-\varepsilon_{v'w'}\right)}\\
&  & \times\left(1+\left[v\leftrightarrow w,\,v'\leftrightarrow w'\right]\right)+\text{c.c.}\\
D_{2} & = & {\sum_{abmn}}'\frac{g_{abmn}g_{mnvw}\tilde{g}_{v'w'ab}}{\left(\varepsilon_{abvw}-\varepsilon_{mnv'w'}\right)}\bigg[\frac{1}{\left(\varepsilon_{vw}-\varepsilon_{mn}\right)}\\
&  & \mbox{}+\frac{1}{\left(\varepsilon_{ab}-\varepsilon_{v'w'}\right)}\bigg]+\text{c.c.} \\
D_{3} & = & -{\sum_{abmn}}'\frac{\tilde{g}_{abmn}g_{w'nab}\tilde{g}_{v'mvw}}{\left(\varepsilon_{abvw}-\varepsilon_{mnv'w'}\right)}\bigg[\frac{1}{\left(\varepsilon_{ab}-\varepsilon_{nw'}\right)}\\
&  & \mbox{}+\frac{1}{\left(\varepsilon_{vw}-\varepsilon_{mv'}\right)}\bigg]
\left(1+\left[v\leftrightarrow w,\,v'\leftrightarrow w'\right]\right)+\text{c.c.} \\
D_{4} & = & \sum_{abmn}\frac{\tilde{g}_{abmn}\tilde{g}_{w'nwb}\tilde{g}_{v'mva}}{\left(\varepsilon_{bw}-\varepsilon_{nw'}\right)\left(\varepsilon_{av}-\varepsilon_{mv'}\right)}(1-[v\leftrightarrow w])+\text{c.c.}\\
D_{5} & = & -\sum_{abmn}\frac{\tilde{g}_{w'amn}\tilde{g}_{nbaw}\tilde{g}_{v'mvb}}{\left(\varepsilon_{avw}-\varepsilon_{mnv'}\right)\left(\varepsilon_{bv}-\varepsilon_{mv'}\right)}\\
&  & \times(1-[v\leftrightarrow w])\left(1-\left[v'\leftrightarrow w'\right]\right)+\text{c.c.}\\
D_{6} & = & \sum_{abmn}\frac{\tilde{g}_{w'amn}\tilde{g}_{nbvw}\tilde{g}_{v'mab}}{\left(\varepsilon_{avw}-\varepsilon_{mnv'}\right)\left(\varepsilon_{ab}-\varepsilon_{mv'}\right)}\\
&  & \times\left(1+\left[v\leftrightarrow w,\,v'\leftrightarrow w'\right]\right)+\text{c.c.}\\
D_{7} & = & -\sum_{abmn}\frac{\tilde{g}_{abmw}\tilde{g}_{w'mbn}\tilde{g}_{v'nva}}{\left(\varepsilon_{abv}-\varepsilon_{mv'w'}\right)\left(\varepsilon_{av}-\varepsilon_{nv'}\right)}\\
&  & \times(1-[v\leftrightarrow w])\left(1-\left[v'\leftrightarrow w'\right]\right)+\text{c.c.}\\
D_{8} & = & \sum_{abmn}\frac{\tilde{g}_{w'bwn}\tilde{g}_{nabm}\tilde{g}_{v'mva}}{\left(\varepsilon_{bv}-\varepsilon_{nv'}\right)\left(\varepsilon_{av}-\varepsilon_{mv'}\right)}\\
&  & \times(1-[v\leftrightarrow w])\left(1-\left[v'\leftrightarrow w'\right]\right)\\
D_{9} & = & -\sum_{abmn}\frac{g_{abwm}\tilde{g}_{v'mvn}\tilde{g}_{w'nab}}{\left(\varepsilon_{abv}-\varepsilon_{mv'w'}\right)\left(\varepsilon_{ab}-\varepsilon_{nw'}\right)}\\
&  & \times(1-[v\leftrightarrow w])\left(1-\left[v'\leftrightarrow w'\right]\right)
\end{eqnarray*}
\begin{eqnarray*}
D_{10} & = & -\sum_{abmn}\frac{g_{w'bmn}\tilde{g}_{v'avb}\tilde{g}_{mnwa}}{\left(\varepsilon_{bvw}-\varepsilon_{mnv'}\right)\left(\varepsilon_{aw}-\varepsilon_{mn}\right)}\\
&  & \times(1-[v\leftrightarrow w])\left(1-\left[v'\leftrightarrow w'\right]\right)
\\
T_{1} & = & {\sum_{amnr}}'\frac{g_{w'anm}\tilde{g}_{mnar}\tilde{g}_{v'rvw}}{\left(\varepsilon_{avw}-\varepsilon_{mnv'}\right)\left(\varepsilon_{vw}-\varepsilon_{rv'}\right)}\\
&  & \times\left(1+\left[v\leftrightarrow w,\,v'\leftrightarrow w'\right]\right)+\text{c.c.}\\
T_{2} & = & \sum_{amnr}\frac{g_{w'amn}\tilde{g}_{mnwr}\tilde{g}_{v'rva}}{\left(\varepsilon_{avw}-\varepsilon_{mnv'}\right)\left(\varepsilon_{av}-\varepsilon_{rv'}\right)}\\
&  & \times(1-[v\leftrightarrow w])\left(1-\left[v'\leftrightarrow w'\right]\right)+\text{c.c.}\\
T_{3} & = & {\sum_{amnr}}'\frac{\tilde{g}_{w'anr}\tilde{g}_{v'rma}\tilde{g}_{mnvw}}{\left(\varepsilon_{avw}-\varepsilon_{nrv'}\right)\left(\varepsilon_{vw}-\varepsilon_{mn}\right)}\\
&  & \times\left(1+\left[v\leftrightarrow w,\, v'\leftrightarrow w'\right]\right)+\text{c.c.}\\
T_{4} & = & \sum_{amnr}\frac{\tilde{g}_{w'arn}\tilde{g}_{v'nvm}\tilde{g}_{rmwa}}{\left(\varepsilon_{avw}-\varepsilon_{nrv'}\right)\left(\varepsilon_{aw}-\varepsilon_{mr}\right)}\\
&  & \times(1-[v\leftrightarrow w])\left(1-\left[v'\leftrightarrow w'\right]\right)\\
Q     & = & {\sum_{mnrs}}'\frac{g_{v'w'rs}g_{rsmn}\tilde{g}_{mnvw}}{\left(\varepsilon_{vw}-\varepsilon_{rs}\right)\left(\varepsilon_{vw}-\varepsilon_{mn}\right)}\\
B_{1} & = & -\sum_{amnx}\frac{\tilde{g}_{w'amn}g_{nmax}\tilde{g}_{v'xvw}}{\left(\varepsilon_{avw}-\varepsilon_{mnv'}\right)\left(\varepsilon_{ax}-\varepsilon_{mn}\right)}\\
&  & \times\left(1+\left[v\leftrightarrow w,\,v'\leftrightarrow w'\right]\right)\\
B_{2} & = & -\sum_{amxy}\frac{\tilde{g}_{v'axm}\tilde{g}_{w'mya}\tilde{g}_{xyvw}}{\left(\varepsilon_{avw}-\varepsilon_{mxw'}\right)\left(\varepsilon_{ay}-\varepsilon_{mw'}\right)}\\
&  & \times\left(1+\left[v\leftrightarrow w,\,v'\leftrightarrow w'\right]\right)\\
B_{3} & = & -{\sum_{mnxy}}'\frac{g_{v'w'mn}g_{mnxy}\tilde{g}_{xyvw}}{\left(\varepsilon_{vw}-\varepsilon_{mn}\right)\left(\varepsilon_{xy}-\varepsilon_{mn}\right)},
\end{eqnarray*}}%
where $g_{ijkl}\equiv\left\langle ij\left|r_{12}^{-1}\right|kl\right\rangle $
is the Coulomb matrix element defined in terms of single-electron functions
as
\[
g_{ijkl}=\int\!\!\!\int\! d\mathbf{r_{1}}d\mathbf{r_{2}}
\frac{1}{\left|\mathbf{r_{1}-r_{2}}\right|}\phi_{i}^{\dagger}\!
\left(\mathbf{r_{1}}\right)\phi_{j}^{\dagger}\!
\left(\mathbf{r_{2}}\right)\phi_{k}\!
\left(\mathbf{r_{1}}\right)\phi_{l}\!
\left(\mathbf{r_{2}}\right),
\]
 and $\tilde{g}_{ijkl}\equiv g_{ijkl}-g_{ijlk}$.
The notation
$\varepsilon_{ijkl}\equiv\varepsilon_{i}+\varepsilon_{j}+\varepsilon_{k}+\varepsilon_{l}$,
etc.\ for the sum of single-electron eigenenergies has also been
used. The third-order terms are arranged by the number of
excited states in the sums over states. Zero, single, double, triple,
quadruple excited-state terms are designated by the letters
$Z,S,D,T,Q$. Terms associated with backwards (folded) diagrams are
designated by $B$. Backwards diagrams are unique for open-shell
systems and exist only in the third or higher order of MBPT. The
summation indices $(a,b,c,d)$ refer to core states, $(m,n,r,s)$
refer to excited states and, in backward diagrams, indices $(x,y)$
refer to valence states. The primes above the summation signs
indicate that excited states $(m,n,r,s)$ are restricted to the
orthogonal space $Q$ only. This restriction applies only to the term with denominator
$1/(\varepsilon_{vw}-\varepsilon_{mn})$ in  $D_2$ and to the term with denominator
$1/(\varepsilon_{vw}-\varepsilon_{mv'})$ in $D_3$.
The c.c.\ denotes complex conjugate. The
conjugate diagrams are obtained by a reflection through a horizontal
axis, with the initial and final states switched
$(vw)\leftrightarrow\left(v^{\prime}w^{\prime}\right)$.
Direct diagrams of the two-body terms are shown in
Fig.~\ref{cap:Third-II}.
Technically, there are subtle changes in energy denominators in going from the term presented to
its c.c.\ counterpart.
These changes can be deduced by re-drawing the diagram upside-down and reading off the new denominators.

In $D_1$ and $D_4$ (but not in $D_2$ and $D_3$) we combined diagrams associated
with double excitations that
have the same numerators but different denominators using the formula
\[
\frac{1}{(A+B)A}+\frac{1}{(A+B)B}=\frac{1}{AB}.
\]
Diagrams $D_1$ -- $D_4$ are special in the sense that two orderings
are possible. The ambiguous vertices are labeled by crosses in the
diagrams. Many of the third-order terms in the two-body part have
external exchanges and complex conjugates so each diagram
illustrated in Fig.~\ref{cap:Third-II} has from one to eight
variants. The largest fraction of computer time is spent evaluating
the term $Q$ and most of the remainder is spent on repetitive
evaluation of terms $D_k$, $T_k$ and their variants. Angular
decompositions of the direct formulas are listed in the Appendix.

\section{Application and discussion}\label{sec:Discussion}

As a first illustration, we apply the theory described above to obtain energies
of the ground state and excited states in the $n=2$ complex
for Be-like ions.
In Table~\ref{cap:Table-Be1}, we give a detailed breakdown of the contributions
from first-, second- and third-order perturbation theory, together with corrections from the
Breit interaction and from the Lamb-shift, for excitation energies of $(2s2p)\, ^3\! P_{0,1,2}$ states
of Be-like ions.
The experimental energies are
taken from the NIST database for atomic spectroscopy~\cite{key-NIST2006}.
Energies $E^{(0+1)}$ represent the lowest-order energies including
the Breit correction. Lowest-order Lamb shifts $E_{\text{Lamb}}$ are
obtained  following the method described in Ref.~\cite{key-Cheng1993}.
We find that the residual differences between calculated and measured energies
$\Delta E$ decrease rapidly  with increasing $Z$.
This is expected since MBPT converges better for charged ions than
for neutral atoms. In fact, for highly-charged ions, correlations
are expected to decrease approximately as $Z^{2-n}$, where $n$ is
the order of perturbation theory~\cite{key-Sapirstein1993}.  On the
other hand, QED effects (Lamb shifts) become more important along
an isoelectronic sequence. The results in Table~\ref{cap:Table-Be1},
confirm both of these trends.
Results of our calculations of excitation energies of all levels in the $n=2$ complex
for Be-like N (N\,IV) are presented in Table~\ref{cap:Table-Be2} and are seen to be in agreement with
measurement to parts in $10^4$.%
\begin{table}
\caption{Comparisons of third-order energies (cm$^{-1}$) of the triplet
  $(2s2p)\,^3P$ states of Be-like ions $Z$=4--7 with measurements are
  given, illustrating the rapid $\left(1/Z^3\right)$ decrease of the
  residual correlation corrections with increasing $Z$. A breakdown of
  contributions to the energy from Coulomb and Breit correlation
  corrections and the Lamb shift is given.}\label{cap:Table-Be1}
\begin{ruledtabular}
\begin{tabular}{lrrrr}
\multicolumn{1}{c}{$Z$} &
\multicolumn{1}{c}{4}   &
\multicolumn{1}{c}{5}   &
\multicolumn{1}{c}{6}   &
\multicolumn{1}{c}{7}\\
\hline
\multicolumn{5}{c}{$(2s2p)\,^{3}P_{0}$}\\
$E^{(0+1)}$      & 23607.9& 39116.7& 54204.5& 69072.2\\
$E^{(2)}$        & -3114.3& -2583.2& -2344.0& -2201.5\\
$B^{(2)}$        &    -1.7&    -3.7&    -6.7&   -10.5\\
$E^{(3)}$        &   473.5&   598.0&   412.9&   294.9\\
$E_{\text{Lamb}}$&    -0.9&    -3.3&    -8.4&   -17.7\\
$E_{\text{tot}}$ & 20964.6& 37124.4& 52258.4& 67137.3\\
$E_{\text{expt}}$& 21978.3& 37336.7& 52367.1& 67209.2\\
$\Delta E$       & -1014  &  -212  &  -109  &   -72\\[0.3ex]
\multicolumn{5}{c}{$(2s2p)\,^{3}P_{1}$}\\
$E^{(0+1)}$      & 23607.4& 39120.2& 54223.4& 69127.7\\
$E^{(2)}$        & -3113.9& -2582.2& -2342.2& -2198.6\\
$B^{(2)}$        &    -0.6&    -1.7&    -3.3&    -5.6\\
$E^{(3)}$        &   473.4&   597.8&   412.7&   294.7\\
$E_{\text{Lamb}}$&    -0.9&    -3.2&    -8.3&   -17.5\\
$E_{\text{tot}}$ & 20965.5& 37130.9& 52282.3& 67200.7\\
$E_{\text{expt}}$& 21978.9& 37342.4& 52390.8& 67272.3\\
$\Delta E$       & -1013  &  -212  &  -109  &   -72\\[0.3ex]
\multicolumn{5}{c}{$(2s2p)\,^{3}P_{2}$}\\
$E^{(0+1)}$      & 23608.2& 39132.6& 54272.9& 69260.8\\
$E^{(2)}$        & -3113.0& -2580.1& -2338.4& -2192.6\\
$B^{(2)}$        &     0.4&     0.5&     0.4&     0.0\\
$E^{(3)}$        &   473.2&   597.4&   412.3&   294.1\\
$E_{\text{Lamb}}$&    -0.8&    -3.2&    -8.2&   -17.1\\
$E_{\text{tot}}$ & 20968.0& 37147.3& 52339.1& 67345.2\\
$E_{\text{expt}}$& 21981.3& 37358.3& 52447.1& 67416.3\\
$\Delta E$       & -1013  &  -211  &  -108  &   -71
\end{tabular}
\end{ruledtabular}
\end{table}%
\begin{table*}
\caption{Third-order energies (cm$^{-1}$) of states in the $n=2$ complex
 of the Be-like ion N\,IV,
 including corrections for the Breit interaction and the Lamb shift.\label{cap:Table-Be2}}
\begin{ruledtabular}
\begin{tabular}{lrrrrrrrrr}
\multicolumn{1}{c}{}          &
\multicolumn{1}{c}{$^3P^o_0$} &
\multicolumn{1}{c}{$^3P^o_1$} &
\multicolumn{1}{c}{$^3P^o_2$} &
\multicolumn{1}{c}{$^1P^o_1$} &
\multicolumn{1}{c}{$^3P^e_0$} &
\multicolumn{1}{c}{$^3P^e_1$} &
\multicolumn{1}{c}{$^3P^e_2$} &
\multicolumn{1}{c}{$^1D^e_2$} &
\multicolumn{1}{c}{$^1S^e_0$}\\
\hline
$E_\text{tot}$ & 67137.3& 67200.7& 67345.2& 130764.1& 175499.4& 175572.8& 175699.0& 188899.9& 235421.9\\
$E_\text{expt}$& 67209.2& 67272.3& 67416.3& 130693.9& 175535.4& 175608.1& 175732.9& 188882.5& 235369.3\\
$\Delta E$     &   -72  &   -72  &   -71  &     70  &    -36  &    -35  &    -34  &     17  &     53
\end{tabular}
\end{ruledtabular}
\end{table*}%
\begin{table}
\caption{Comparison with measurement of theoretical energies (cm$^{-1}$) of some of the low-lying states
in the $n=3$ complex of the Mg-like ion P\,IV, including a breakdown of contributions from
Coulomb and Breit correlation corrections and the Lamb shift.}\label{cap:Table-P4}
\begin{ruledtabular}
\begin{tabular}{lrrr}
\multicolumn{1}{c}{}                            &
\multicolumn{1}{c}{$(3s3p)\,^3P_0$} &
\multicolumn{1}{c}{$(3s3p)\,^3P_1$} &
\multicolumn{1}{c}{$(3s3p)\,^3P_2$}\\
\hline
$E^{(0+1)}$     &  67021.3& 67242.9& 67696.5\\
$E^{(2)}$       &    110.3&   116.0&   130.1\\
$B^{(2)}$       &     -0.9&     0.3&     1.3\\
$E^{(3)}$       &    807.4&   807.6&   807.6\\
$E_\text{Lamb}$ &    -21.1&   -20.9&   -20.5\\
$E_\text{tot}$  &  67917.1& 68146.0& 68615.0\\
$E_\text{expt}$ &  67918.0& 68146.5& 68615.2\\
$\Delta E$      &     -0.9&    -0.5  &  -0.2\\[0.6ex]
\multicolumn{1}{c}{}                &
\multicolumn{1}{c}{$(3s3p)\,^1P_1$} &
\multicolumn{1}{c}{$\left(3p^2\right){^1D_2}$} &
\multicolumn{1}{c}{$\left(3p^2\right){^3P_0}$}\\
\hline
$E^{(0+1)}$     & 120479.5& 180554.7& 165971.6\\
$E^{(2)}$       & -20906.0& -61699.8&  -2027.7\\
$B^{(2)}$       &    -15.8&     -8.4&     -5.2\\
$E^{(3)}$       &   6470.7&  48769.6&   1089.6\\
$E_\text{Lamb}$ &    -20.7&    -43.6&    -44.0\\
$E_\text{tot}$  & 106007.7& 167572.4& 164984.3\\
$E_\text{expt}$ & 105190.4& 166144.0& 164941.4\\
$\Delta E$      &    817  &   1428  &     43\\[0.6ex]
\multicolumn{1}{c}{}                &
\multicolumn{1}{c}{$\left(3p^2\right){^3P_1}$} &
\multicolumn{1}{c}{$\left(3p^2\right){^3P_2}$} &
\multicolumn{1}{c}{$\left(3p^2\right){^1S_0}$}\\
\hline
$E^{(0+1)}$     & 166200.8& 166633.3& 212201.4\\
$E^{(2)}$       &  -2013.9&  -2008.1& -23060.7\\
$B^{(2)}$       &     -4.8&     -2.7&    -24.6\\
$E^{(3)}$       &   1087.9&   1077.3&   5810.7\\
$E_\text{Lamb}$ &    -43.8&    -43.4&    -41.3\\
$E_\text{tot}$  & 165226.1& 165656.5& 194885.6\\
$E_\text{expt}$ & 165185.4& 165654.0& 194591.8\\
$\Delta E$      &     41  &      3  &    294\\
\end{tabular}
\end{ruledtabular}
\end{table}

As a more involved example, we give a complete breakdown of
contributions to energies of low-lying states in the $n=3$
complex for the Mg-like ion P\,IV in Table~\ref{cap:Table-P4}.
For both N\,IV and P\,IV, correlations are seen to account for
about 10\% of the total energies.
For Be-like ions, the third-order
correlations are only 7--10\% of the second-order correlations.
The second-order correlation energies are an order of magnitude smaller
than the corresponding DHF energies. The Breit correction $B^{(2)}$, which is obtained
by linearizing the second-order matrix elements in Breit interaction,
is also small for such lightly-charged ions.

Our calculations for Be-like ions are able to produce results accurate
to order of ten cm$^{-1}$. This shows that the third-order energy
correction is very important for divalent ions. By comparison, the
second-order results of \citet{key-Safronova1996} for N\,IV agree with experiment
at the level of a few hundred cm$^{-1}$.
It should be noted that results from the CI+MBPT method~\cite{key-Savukov2002}
mentioned in the introduction are consistently more accurate
than the present third-order results. However, the
CI+MBPT calculations contain a free parameter in the energy denominators
that is adjusted to give optimized energies.
In contrast, third-order MBPT is completely \emph{ab initio}.
For the C\,III ion, the large-scale CI calculations of \citet{key-Chen2001}
mentioned in the introduction also give transition energies accurate to
better than a hundred cm$^{-1}$ on average.  Those large-scale CI
calculations are also \emph{ab initio} and have about the same accuracy
as third-order MBPT for states in C\,III.

For the P\,IV ion, our results are in good agreement with experiment,
the average discrepancy being several hundred cm$^{-1}$.
\citet{key-Chaudhuri1998} employed
an \emph{effective valence shell Hamiltonian} (EVSH) to calculate
energies of Mg-like ions and obtained results for P\,IV having a
discrepancy of about a thousand cm$^{-1}$, which is somewhat larger, but
comparable, to the accuracy of our third-order calculations.

It is informative to analyze our results in terms of diagrams. The
relative contributions of the third-order two-body terms for the
ground-state energy of a typical member in the Be sequence and the
ion P\,IV are summarized in Tables~\ref{cap:Table-C3-rel} and
\ref{cap:Table-P4-rel}. Dominant classes of diagrams are $Q$ and
$B_{3}$. This is understandable since $Q$ are quadruple excited-state
diagrams and involve no core excitations. The quadruple excitation
diagrams are entirely due to valence-valence correlation effects;
they are expected to be large because of the strong repulsion of
the outer valence electrons. Class $B_{3}$ are backwards diagrams,
which are characteristic of open-shell systems. As shown in
Fig.~\ref{cap:Third-II}, this class is also associated solely with
valence-valence correlation. The two classes $Q$ and $B_{3}$ tend
to cancel each other as there is an extra phase of -1 associated
with backwards diagrams. It is interesting to note that even after
subtraction of the contributions from $Q$ and $B_{3}$, their
difference is still larger
than the contribution from any other single class of diagrams for C\,III.%
\begin{table}
\caption{Relative contributions of third-order two-body terms for C\,III.}\label{cap:Table-C3-rel}
\begin{ruledtabular}
\begin{tabular}{lrlrlrlrlr}
\multicolumn{1}{c}{Term} &
\multicolumn{1}{c}{(\%)}   &
\multicolumn{1}{c}{Term} &
\multicolumn{1}{c}{(\%)}   &
\multicolumn{1}{c}{Term} &
\multicolumn{1}{c}{(\%)}   &
\multicolumn{1}{c}{Term} &
\multicolumn{1}{c}{(\%)}   &
\multicolumn{1}{c}{Term} &
\multicolumn{1}{c}{(\%)}\\
\hline
$Z$    &  0.1& $D_{1}$& -0.2&  $D_{6}$&  0.0& $T_{1}$&  4.2& $B_{1}$& -0.8\\
$S_{1}$&  0.1& $D_{2}$&  0.4&  $D_{7}$&  0.0& $T_{2}$& -0.3& $B_{2}$& -0.1\\
$S_{2}$& -0.4& $D_{3}$& -1.6&  $D_{8}$&  0.3& $T_{3}$& -0.5& $B_{3}$& -46.1\\
$S_{3}$& -0.1& $D_{4}$&  0.1&  $D_{9}$& -0.1& $T_{4}$&  2.0\\
$S_{4}$&  0.4& $D_{5}$&  0.2& $D_{10}$& -1.0&     $Q$& 41.2\\
\end{tabular}
\end{ruledtabular}
\end{table}%
\begin{table}
\caption{Relative contributions of third-order two-body terms for P\,IV.}\label{cap:Table-P4-rel}
\begin{ruledtabular}
\begin{tabular}{lrlrlrlrlr}
\multicolumn{1}{c}{Term} &
\multicolumn{1}{c}{(\%)}   &
\multicolumn{1}{c}{Term} &
\multicolumn{1}{c}{(\%)}   &
\multicolumn{1}{c}{Term} &
\multicolumn{1}{c}{(\%)}   &
\multicolumn{1}{c}{Term} &
\multicolumn{1}{c}{(\%)}   &
\multicolumn{1}{c}{Term} &
\multicolumn{1}{c}{(\%)}\\
\hline
$Z$    &  0.1& $D_{1}$& -0.4&  $D_{6}$& -0.1& $T_{1}$&  9.4& $B_{1}$& -3.9\\
$S_{1}$&  0.4& $D_{2}$&  0.3&  $D_{7}$& -0.3& $T_{2}$&  0.1& $B_{2}$& -0.7\\
$S_{2}$& -0.3& $D_{3}$& -3.6&  $D_{8}$&  0.1& $T_{3}$&  2.1& $B_{3}$& -29.2\\
$S_{3}$&  0.2& $D_{4}$&  0.6&  $D_{9}$& -0.5& $T_{4}$& 10.2\\
$S_{4}$&  1.4& $D_{5}$& -1.1& $D_{10}$& -5.0&     $Q$& 29.9\\
\end{tabular}
\end{ruledtabular}
\end{table}

\section{Conclusions}

The accuracy of third-order MBPT results is at 0.2\% level for
lightly-charged ions of both Be and Mg isoelectronic sequences. This level of accuracy
is comparable or superior to the two \emph{ab initio} methods mentioned
in Sec.~\ref{sec:Discussion}. A complete third-order calculation is
important to understand the relative importance of different
contributions to energies of divalent systems. The folded diagrams as
well as the quadruple excitation diagrams are significant. The dominant
role of these two classes of diagrams is attributed to the strong
correlation of the two valence electrons. This conclusion is useful for
workers developing combined CI-MBPT methods which include dominant
third-order diagrams. It is also helpful for researchers setting up SDCC
calculations as they try to classify and account for the contributions
from the third-order diagrams associated with omitted triple
excitations. Although one might expect a complete fourth-order
calculation for divalent systems to improve the accuracy of the present
calculations still further, it is unlikely that such a complex
calculation will be carried out in the near future.

\begin{acknowledgments}
The work of H.C.H.\ and W.R.J.\ was supported in part by National
Science Foundation (NSF) Grant No. PHY-04-56828. The work of M.S.S.\ was
supported in part by NSF Grant No. PHY-04-57078.
\end{acknowledgments}

\appendix

\section*{Appendix}

Angular decompositions of direct formulas for the third-order two-body
part are presented.%
{\small\begin{eqnarray*}
Z & = & \sum_{\stackrel{L_{1}L_{2}L_{3}}{{ {a}{b}{c}{d}}}}\frac{X_{L_{1}}(cdwv)X_{L_{2}}(abcd)Z_{L_{3}}\left(w'v'ab\right)}{\left(\varepsilon_{ab}-\varepsilon_{v'w'}\right)\left(\varepsilon_{cd}-\varepsilon_{v'w'}\right)}\\
&  & \times(-1)^{J+L_{1}+L_{2}+L_{3}+j_{a}+j_{b}+j_{c}+j_{d}+j_{w'}+j_{v}}\\
&  & \times\left\{\begin{array}{rrr}
J & j_{c} & j_{d}\\
L_{1} & j_{v} & j_{w}\end{array}\right\}%
\left\{\begin{array}{rrr}
j_{a} & j_{b} & J\\
j_{d} & j_{c} & L_{2}\end{array}\right\}%
\left\{\begin{array}{rrr}
J & j_{a} & j_{b}\\
L_{3} & j_{v'} & j_{w'}\end{array}\right\}\\
S_{1} & = & \frac{1}{\left[j_{w}\right]}{\sum_{\stackrel{LL'}{{{a}{b}{c}{m}}}}}\frac{X_{L}(acmw)Z_{L}(mbac)Z_{L'}\left(v'w'vb\right)}{\left(\varepsilon_{bv}-\varepsilon_{v'w'}\right)\left(\varepsilon_{acv}-\varepsilon_{mv'w'}\right)}\\
&  & \times(-1)^{J+L'+j_{a}+j_{b}+j_{c}+j_{m}+j_{w'}+j_{v}}\\
&  & \times\delta_{j_{b}j_{w}}\frac{1}{[L]}%
\left\{\begin{array}{rrr}
J & j_{v'} & j_{w'}\\
L' & j_{b} & j_{v}\end{array}\right\}\\
S_{2} & = & \sum_{\stackrel{LL'}{{{a}{b}{c}{m}}}}\frac{Z_{L}(abwm)Z_{L}(mcbv)Z_{L'}\left(v'w'ca\right)}{\left(\varepsilon_{ac}-\varepsilon_{v'w'}\right)\left(\varepsilon_{abv}-\varepsilon_{mv'w'}\right)}\\
&  & \times(-1)^{1+L'+j_{a}+j_{b}+j_{c}+j_{m}+j_{w'}+j_{v}}\\
&  & \times\frac{1}{[L]}
\left\{\begin{array}{rrr}
J & j_{a} & j_{c}\\
L & j_{v} & j_{w}\end{array}\right\}%
\left\{\begin{array}{rrr}
J & j_{c} & j_{a}\\
L' & j_{w'} & j_{v'}\end{array}\right\}
\\
S_{3} & = & \sum_{\stackrel{L_{1}L_{2}L_{3}}{{{a}{b}{c}{m}}}}\frac{X_{L_{1}}(abwm)Z_{L_{2}}\left(w'cab\right)Z_{L_{3}}\left(v'mvc\right)}{\left(\varepsilon_{cv}-\varepsilon_{mv'}\right)\left(\varepsilon_{abv}-\varepsilon_{mv'w'}\right)}\\
 &  & \times(-1)^{J+j_{v}+j_{w}}%
\left\{\begin{array}{rrr}
L_{1} & L_{2} & L_{3}\\
j_{c} & j_{m} & j_{b}\end{array}\right\}\\
&  & \times\left\{\begin{array}{rrr}
j_{w} & j_{w'} & L_{3}\\
L_{2} & L_{1} & j_{a}\end{array}\right\}%
\left\{\begin{array}{rrr}
J & j_{v'} & j_{w'}\\
L_{3} & j_{w} & j_{v}\end{array}\right\}\\
S_{4} & = & \sum_{\stackrel{L_{1}L_{2}L_{3}}{{{a}{b}{c}{m}}}}\frac{Z_{L_{1}}(abwm)Z_{L_{2}}\left(v'cvb\right)Z_{L_{3}}\left(w'mac\right)}{\left(\varepsilon_{ac}-\varepsilon_{mw'}\right)\left(\varepsilon_{abv}-\varepsilon_{mv'w'}\right)}\\
&  & \times(-1)^{J+L_{1}+L_{2}+L_{3}+j_{v}+j_{w}}%
\left\{\begin{array}{rrr}
J & j_{v'} & j_{w'}\\
L_{2} & j_{w} & j_{v}\end{array}\right\}\\
&  & \times\left\{\begin{array}{rrr}
j_{w} & j_{w'} & L_{2}\\
L_{3} & L_{1} & j_{a}\end{array}\right\}%
\left\{\begin{array}{rrr}
L_{1} & L_{2} & L_{3}\\
j_{c} & j_{m} & j_{b}\end{array}\right\} \\
D_{1} & = & -\frac{1}{\left[j_{w}\right]}\sum_{\stackrel{LL'}{{{a}{b}{m}{n}}}}\frac{X_{L}(abmn)Z_{L}(mnwb)Z_{L'}\left(v'w'va\right)}{\left(\varepsilon_{bw}-\varepsilon_{mn}\right)\left(\varepsilon_{av}-\varepsilon_{v'w'}\right)}\\
&  & \times(-1)^{J+L'+j_{a}+j_{b}+j_{m}+j_{n}+j_{w'}+j_{v}}\\
&  & \times\delta_{j_{a}j_{w}}\frac{1}{[L]}%
\left\{\begin{array}{rrr}
J & j_{v'} & j_{w'}\\
L' & j_{a} & j_{v}\end{array}\right\}
\end{eqnarray*}
\begin{eqnarray*}
D_{2} & = & {\sum_{\stackrel{L_{1}L_{2}L_{3}}{{{a}{b}{m}{n}}}}} '\  \frac{X_{L_{1}}(abmn)X_{L_{2}}(mnvw)Z_{L_{3}}\left(v'w'ab\right)}{\left(\varepsilon_{abvw}-\varepsilon_{mnv'w'}\right)}\\
&  & \times\left[\frac{1}{\left(\varepsilon_{vw}-\varepsilon_{mn}\right)}+\frac{1}{\left(\varepsilon_{ab}-\varepsilon_{v'w'}\right)}\right]\\
&  & \times(-1)^{J+L_{1}+L_{2}+L_{3}+j_{a}+j_{b}+j_{m}+j_{n}+j_{w'}+j_{v}}\\
&  & \times\left\{\begin{array}{rrr}
j_{a} & j_{b} & J\\
j_{n} & j_{m} & L_{1}\end{array}\right\}%
\left\{\begin{array}{rrr}
J & j_{m} & j_{n}\\
L_{2} & j_{w} & j_{v}\end{array}\right\}%
\left\{\begin{array}{rrr}
J & j_{a} & j_{b}\\
L_{3} & j_{w'} & j_{v'}\end{array}\right\}\\
D_{3} & = & -\frac{1}{\left[j_{w'}\right]}{\sum_{\stackrel{LL'}{{{a}{b}{m}{n}}}}}'\ \frac{Z_{L}(abmn)X_{L}\left(w'nab\right)Z_{L'}\left(v'mvw\right)}{\left(\varepsilon_{abvw}-\varepsilon_{mnv'w'}\right)}\\
&  & \times\left[\frac{1}{\left(\varepsilon_{ab}-\varepsilon_{nw'}\right)}+\frac{1}{\left(\varepsilon_{vw}-\varepsilon_{mv'}\right)}\right]\\
&  & \times(-1)^{J+L'+j_{a}+j_{b}+j_{m}+j_{n}+j_{w'}+j_{v}}\\
&  & \times\delta_{j_{m}j_{w'}}\frac{1}{[L]}%
\left\{\begin{array}{rrr}
J & j_{v'} & j_{m}\\
L' & j_{w} & j_{v}\end{array}\right\}\\
D_{4} & = & {\displaystyle \sum_{\stackrel{L}{{{a}{b}{m}{n}}}}}\frac{Z_{L}(abmn)Z_{L}\left(w'nwb\right)Z_{L}\left(v'mva\right)}{\left(\varepsilon_{bw}-\varepsilon_{nw'}\right)\left(\varepsilon_{av}-\varepsilon_{mv'}\right)}\\
&  & \times(-1)^{J+L+j_{a}+j_{b}+j_{m}+j_{n}+j_{w'}+j_{v}}\\
&  & \times\frac{1}{[L]^{2}}%
\left\{\begin{array}{rrr}
J & j_{w'} & j_{v'}\\
L & j_{v} & j_{w}\end{array}\right\}
\\
D_{5} & = & -\sum_{\stackrel{LL'}{{{a}{b}{m}{n}}}}\frac{Z_{L}\left(w'amn\right)Z_{L}(nbaw)Z_{L'}\left(v'mvb\right)}{\left(\varepsilon_{avw}-\varepsilon_{mnv'}\right)\left(\varepsilon_{bv}-\varepsilon_{mv'}\right)}\\
&  & \times(-1)^{J+1+L+j_{a}+j_{b}+j_{m}+j_{n}+j_{w'}+j_{v}}\\
&  & \times\frac{1}{[L]}%
\left\{\begin{array}{rrr}
J & j_{v'} & j_{w'}\\
L' & j_{w} & j_{v}\end{array}\right\}%
\left\{\begin{array}{rrr}
j_{w'} & j_{w} & L'\\
j_{b} & j_{m} & L\end{array}\right\}\\
D_{6} & = & \sum_{\stackrel{L_{1}L_{2}L_{3}}{{{a}{b}{m}{n}}}}\frac{Z_{L_{1}}\left(w'amn\right)Z_{L_{2}}(nbvw)Z_{L_{3}}\left(v'mab\right)}{\left(\varepsilon_{avw}-\varepsilon_{mnv'}\right)\left(\varepsilon_{ab}-\varepsilon_{mv'}\right)}\\
&  & \times(-1)^{1+L_{1}+L_{2}+L_{3}+j_{v'}+j_{v}}\\
&  & \times\left\{\begin{array}{rrr}
J & j_{n} & j_{b}\\
L_{2} & j_{w} & j_{v}\end{array}\right\}%
\left\{\begin{array}{ccc}
J & j_{n} & j_{b}\\
j_{w'} & L_{1} & j_{m}\\
j_{v'} & j_{a} & L_{3}\end{array}\right\}\\
D_{7} & = & -\sum_{\stackrel{L_{1}L_{2}L_{3}}{{{a}{b}{m}{n}}}}\frac{Z_{L_{1}}(abmw)Z_{L_{2}}\left(w'mbn\right)Z_{L_{3}}\left(v'nva\right)}{\left(\varepsilon_{abv}-\varepsilon_{mv'w'}\right)\left(\varepsilon_{av}-\varepsilon_{nv'}\right)}\\
&  & \times(-1)^{J+L_{1}+L_{2}+L_{3}+j_{v}+j_{w}}%
\left\{\begin{array}{rrr}
L_{1} & L_{2} & L_{3}\\
j_{n} & j_{a} & j_{m}\end{array}\right\}\\
&  & \times\left\{\begin{array}{rrr}
j_{w} & j_{w'} & L_{3}\\
L_{2} & L_{1} & j_{b}\end{array}\right\}%
\left\{\begin{array}{rrr}
J & j_{v'} & j_{w'}\\
L_{3} & j_{w} & j_{v}\end{array}\right\}
\\
D_{8} & = & \sum_{\stackrel{L}{{{a}{b}{m}{n}}}}\frac{Z_{L}\left(w'bwn\right)Z_{L}(nabm)Z_{L}\left(v'mva\right)}{\left(\varepsilon_{bv}-\varepsilon_{nv'}\right)\left(\varepsilon_{av}-\varepsilon_{mv'}\right)}\\
&  & \times(-1)^{J+L+j_{a}+j_{b}+j_{m}+j_{n}+j_{w'}+j_{v}}\\
&  & \times\frac{1}{[L]^{2}}%
\left\{\begin{array}{rrr}
J & j_{w'} & j_{v'}\\
L & j_{v} & j_{w}\end{array}\right\}\\
\end{eqnarray*}
\begin{eqnarray*}
D_{9} & = & -\sum_{\stackrel{L_{1}L_{2}L_{3}}{{{a}{b}{m}{n}}}}\frac{X_{L_{1}}(abwm)Z_{L_{2}}\left(v'mvn\right)Z_{L_{3}}\left(w'nab\right)}{\left(\varepsilon_{abv}-\varepsilon_{mv'w'}\right)\left(\varepsilon_{ab}-\varepsilon_{nw'}\right)}\\
&  & \times(-1)^{J+j_{v}+j_{w}}%
\left\{\begin{array}{rrr}
L_{1} & L_{2} & L_{3}\\
j_{n} & j_{b} & j_{m}\end{array}\right\}\\
&  & \times\left\{\begin{array}{rrr}
j_{w} & j_{w'} & L_{2}\\
L_{3} & L_{1} & j_{a}\end{array}\right\}%
\left\{\begin{array}{rrr}
J & j_{v'} & j_{w'}\\
L_{2} & j_{w} & j_{v}\end{array}\right\}\\
D_{10} & = & -\sum_{\stackrel{L_{1}L_{2}L_{3}}{{{a}{b}{m}{n}}}}\frac{X_{L_{1}}\left(w'bmn\right)Z_{L_{2}}\left(v'avb\right)Z_{L_{3}}(mnwa)}{\left(\varepsilon_{bvw}-\varepsilon_{mnv'}\right)\left(\varepsilon_{aw}-\varepsilon_{mn}\right)}\\
&  & \times(-1)^{J+j_{v}+j_{w}}%
\left\{\begin{array}{rrr}
L_{1} & L_{2} & L_{3}\\
j_{a} & j_{n} & j_{b}\end{array}\right\} \\
&  & \times\left\{\begin{array}{rrr}
j_{w'} & j_{w} & L_{2}\\
L_{3} & L_{1} & j_{m}\end{array}\right\}%
\left\{\begin{array}{rrr}
J & j_{v'} & j_{w'}\\
L_{2} & j_{w} & j_{v}\end{array}\right\}
\\
T_{1} & = & \frac{1}{\left[j_{w'}\right]}{\sum_{\stackrel{LL'}{{{a}{m}{n}{r}}}}}'\ \frac{X_{L}\left(w'anm\right)Z_{L}(mnar)Z_{L'}\left(v'rvw\right)}{\left(\varepsilon_{avw}-\varepsilon_{mnv'}\right)\left(\varepsilon_{vw}-\varepsilon_{rv'}\right)}\\
&  & \times(-1)^{J+1+L'+j_{a}+j_{m}+j_{n}+j_{v}}\\
&  & \times\delta_{j_{r}j_{w'}}\frac{1}{[L]}%
\left\{\begin{array}{rrr}
J & j_{v'} & j_{w'}\\
L' & j_{w} & j_{v}\end{array}\right\}\\
T_{2} & = & \sum_{\stackrel{L_{1}L_{2}L_{3}}{{{a}{m}{n}{r}}}}\frac{X_{L_{1}}\left(w'amn\right)Z_{L_{2}}(mnwr)Z_{L_{3}}\left(v'rva\right)}{\left(\varepsilon_{avw}-\varepsilon_{mnv'}\right)\left(\varepsilon_{av}-\varepsilon_{rv'}\right)}\\
&  & \times(-1)^{J+j_{v}+j_{w}}%
\left\{\begin{array}{rrr}
L_{1} & L_{2} & L_{3}\\
j_{r} & j_{a} & j_{n}\end{array}\right\}\\
&  & \times\left\{\begin{array}{rrr}
j_{w'} & j_{w} & L_{3}\\
L_{2} & L_{1} & j_{m}\end{array}\right\}%
\left\{\begin{array}{rrr}
J & j_{v'} & j_{w'}\\
L_{3} & j_{w} & j_{v}\end{array}\right\}
\\
T_{3} & = & {\sum_{\stackrel{LL'}{{{a}{m}{n}{r}}}}}'\ \frac{Z_{L}\left(w'anr\right)Z_{L}\left(v'rma\right)Z_{L'}(mnvw)}{\left(\varepsilon_{avw}-\varepsilon_{nrv'}\right)\left(\varepsilon_{vw}-\varepsilon_{mn}\right)}\\
&  & \times(-1)^{1+L'+j_{a}+j_{m}+j_{n}+j_{r}+j_{w'}+j_{v}}\\
&  & \times\frac{1}{[L]}%
\left\{\begin{array}{rrr}
J & j_{n} & j_{m}\\
L & j_{v'} & j_{w'}\end{array}\right\}%
\left\{\begin{array}{rrr}
J & j_{m} & j_{n}\\
L' & j_{w} & j_{v}\end{array}\right\}\\
T_{4} & = & \sum_{\stackrel{L_{1}L_{2}L_{3}}{{{a}{m}{n}{r}}}}\frac{Z_{L_{1}}\left(w'arn\right)Z_{L_{2}}\left(v'nvm\right)Z_{L_{3}}(rmwa)}{\left(\varepsilon_{avw}-\varepsilon_{nrv'}\right)\left(\varepsilon_{aw}-\varepsilon_{mr}\right)}\\
&  & \times(-1)^{J+L_{1}+L_{2}+L_{3}+j_{v}+j_{w}}%
\left\{\begin{array}{rrr}
L_{1} & L_{2} & L_{3}\\
j_{m} & j_{a} & j_{n}\end{array}\right\}\\
&  & \times\left\{\begin{array}{rrr}
j_{w'} & j_{w} & L_{2}\\
L_{3} & L_{1} & j_{r}\end{array}\right\}%
\left\{\begin{array}{rrr}
J & j_{v'} & j_{w'}\\
L_{2} & j_{w} & j_{v}\end{array}\right\}\\
Q & = & {\sum_{\stackrel{L_{1}L_{2}L_{3}}{{{m}{n}{r}{s}}}}}'\ \frac{X_{L_{1}}\left(v'w'rs\right)X_{L_{2}}(rsmn)Z_{L_{3}}(mnvw)}{\left(\varepsilon_{vw}-\varepsilon_{rs}\right)\left(\varepsilon_{vw}-\varepsilon_{mn}\right)}\\
&  & \times(-1)^{J+L_{1}+L_{2}+L_{3}+j_{m}+j_{n}+j_{r}+j_{s}+j_{w'}+j_{v}}\\
&  & \times\left\{\begin{array}{rrr}
J & j_{r} & j_{s}\\
L_{1} & j_{w'} & j_{v'}\end{array}\right\}%
\left\{\begin{array}{rrr}
j_{r} & j_{s} & J\\
j_{n} & j_{m} & L_{2}\end{array}\right\}%
\left\{\begin{array}{rrr}
J & j_{m} & j_{n}\\
L_{3} & j_{w} & j_{v}\end{array}\right\}
\end{eqnarray*}
\begin{eqnarray*}
B_{1} & = & -\frac{1}{\left[j_{w'}\right]}{\displaystyle \sum_{\stackrel{LL'}{{{a}{m}{n}{x}}}}}\frac{Z_{L}\left(w'amn\right)X_{L}(nmax)Z_{L'}\left(v'xvw\right)}{\left(\varepsilon_{avw}-\varepsilon_{mnv'}\right)\left(\varepsilon_{ax}-\varepsilon_{mn}\right)}\\
&  & \times(-1)^{J+1+L'+j_{a}+j_{m}+j_{n}+j_{v}}\\
&  & \times\delta_{j_{x}j_{w'}}\frac{1}{[L]}%
\left\{\begin{array}{rrr}
J & j_{v'} & j_{w'}\\
L' & j_{w} & j_{v}\end{array}\right\}\end{eqnarray*}
\begin{eqnarray*}
B_{2} & = & -\sum_{\stackrel{LL'}{{{a}{m}{x}{y}}}}\frac{Z_{L}\left(v'axm\right)Z_{L}\left(w'mya\right)Z_{L'}(xyvw)}{\left(\varepsilon_{avw}-\varepsilon_{mxw'}\right)\left(\varepsilon_{ay}-\varepsilon_{mw'}\right)}\\
&  & \times(-1)^{1+L'+j_{a}+j_{m}+j_{x}+j_{y}+j_{w'}+j_{v}}\\
&  & \times\frac{1}{[L]}%
\left\{\begin{array}{rrr}
J & j_{x} & j_{y}\\
L & j_{w'} & j_{v'}\end{array}\right\}%
\left\{\begin{array}{rrr}
J & j_{x} & j_{y}\\
L' & j_{w} & j_{v}\end{array}\right\}\\
B_{3} & = & -{\sum_{\stackrel{L_{1}L_{2}L_{3}}{{{m}{n}{x}{y}}}}}'\ \frac{X_{L_{1}}\left(v'w'mn\right)X_{L_{2}}(mnxy)Z_{L_{3}}(xyvw)}{\left(\varepsilon_{vw}-\varepsilon_{mn}\right)\left(\varepsilon_{xy}-\varepsilon_{mn}\right)}\\
&  & \times(-1)^{J+L_{1}+L_{2}+L_{3}+j_{m}+j_{n}+j_{x}+j_{y}+j_{w'}+j_{v}}\\
&  & \times\left\{\begin{array}{rrr}
J & j_{m} & j_{n}\\
L_{1} & j_{w'} & j_{v'}\end{array}\right\}%
\left\{\begin{array}{rrr}
j_{m} & j_{n} & J\\
j_{y} & j_{x} & L_{2}\end{array}\right\}%
\left\{\begin{array}{rrr}
J & j_{x} & j_{y}\\
L_{3} & j_{w} & j_{v}\end{array}\right\}.
\end{eqnarray*}}%
The \emph{effective interaction strength}\[
X_{L}(ijkl)=(-1)^{L}\left\langle i\left\Vert C^{L}\right\Vert k\right\rangle \left\langle j\left\Vert C^{L}\right\Vert l\right\rangle R_{L}(ijkl),\]
 is independent of magnetic quantum numbers. The reduced matrix element
of the $\mathbf{C}^{L}$ tensor is\[
\left\langle i\left\Vert C^{L}\right\Vert k\right\rangle =(-1)^{j_{i}+\frac{1}{2}}\sqrt{\left[j_{i}\right]\left[j_{k}\right]}\left(\begin{array}{ccc}
j_{i} & j_{k} & L\\
-\frac{1}{2} & \frac{1}{2} & 0\end{array}\right)\Pi^{e}\left(l_{i},l_{k},L\right),\]
 where $\left[j_{i}\right]\equiv2j_{i}+1$ is the occupation number
of shell $i$, and\[
\Pi^{e}\left(l_{i},l_{k},L\right)=\left\{ \begin{array}{cl}
1 & \text{if }l_{i}+l_{k}+L\text{ is even}\\
0 & \text{if }l_{i}+l_{k}+L\text{ is odd}\end{array}\right..\]
 The Slater integral $R_{L}(ijkl)$ is
 \begin{multline*}
R_{L}(ijkl)  =  \int_{0}^{\infty}\!\int_{0}^{\infty}\! dr_{1}dr_{2}\,\frac{r_{<}^{L}}{r_{>}^{L+1}} \\
\left[P_{i}\left(r_{1}\right)\!P_{k}\left(r_{1}\right)+Q_{i}\left(r_{1}\right)\!Q_{k}\left(r_{1}\right)\right]\\
\times\left[P_{j}\left(r_{2}\right)\!P_{l}\left(r_{2}\right)+Q_{j}\left(r_{2}\right)\!Q_{l}\left(r_{2}\right)\right].
\end{multline*}
 The quantity $Z_{L}(ijkl)$ is defined by
\[
Z_{L}(ijkl)\equiv X_{L}(ijkl)+[L]\sum_{L'}\left\{ \begin{array}{ccc}
j_{j} & j_{l} & L\\
j_{i} & j_{k} & L'\end{array}\right\}\!X_{L'}(ijlk).
\]


\begin{thebibliography}{25}
\expandafter\ifx\csname natexlab\endcsname\relax\def\natexlab#1{#1}\fi
\expandafter\ifx\csname bibnamefont\endcsname\relax
  \def\bibnamefont#1{#1}\fi
\expandafter\ifx\csname bibfnamefont\endcsname\relax
  \def\bibfnamefont#1{#1}\fi
\expandafter\ifx\csname citenamefont\endcsname\relax
  \def\citenamefont#1{#1}\fi
\expandafter\ifx\csname url\endcsname\relax
  \def\url#1{\texttt{#1}}\fi
\expandafter\ifx\csname urlprefix\endcsname\relax\def\urlprefix{URL }\fi
\providecommand{\bibinfo}[2]{#2}
\providecommand{\eprint}[2][]{\url{#2}}

\bibitem[{\citenamefont{Ivanova et~al.}(1967)\citenamefont{Ivanova, Safronova,
  and Tolmachev}}]{key-Ivanova1967}
\bibinfo{author}{\bibfnamefont{A.}~\bibnamefont{Ivanova}},
  \bibinfo{author}{\bibfnamefont{U.}~\bibnamefont{Safronova}},
  \bibnamefont{and}
  \bibinfo{author}{\bibfnamefont{V.}~\bibnamefont{Tolmachev}},
  \bibinfo{journal}{Litov. Phys. Sb.} \textbf{\bibinfo{volume}{7}},
  \bibinfo{pages}{571} (\bibinfo{year}{1967}).

\bibitem[{\citenamefont{Safronova and Ivanova}(1969)}]{key-Safronova1969}
\bibinfo{author}{\bibfnamefont{U.}~\bibnamefont{Safronova}} \bibnamefont{and}
  \bibinfo{author}{\bibfnamefont{A.}~\bibnamefont{Ivanova}},
  \bibinfo{journal}{Opt. Spectrosc} \textbf{\bibinfo{volume}{27}},
  \bibinfo{pages}{193} (\bibinfo{year}{1969}).

\bibitem[{\citenamefont{Ivanova and Safronova}(1975)}]{key-Ivanova1975}
\bibinfo{author}{\bibfnamefont{E.}~\bibnamefont{Ivanova}} \bibnamefont{and}
  \bibinfo{author}{\bibfnamefont{U.}~\bibnamefont{Safronova}},
  \bibinfo{journal}{J. Phys. B} \textbf{\bibinfo{volume}{8}},
  \bibinfo{pages}{1591} (\bibinfo{year}{1975}).

\bibitem[{\citenamefont{Chang}(1989)}]{key-Chang1989}
\bibinfo{author}{\bibfnamefont{T.~N.} \bibnamefont{Chang}},
  \bibinfo{journal}{Phys. Rev. A} \textbf{\bibinfo{volume}{39}},
  \bibinfo{pages}{4946} (\bibinfo{year}{1989}).

\bibitem[{\citenamefont{Chang}(1986)}]{key-Chang1986}
\bibinfo{author}{\bibfnamefont{T.~N.} \bibnamefont{Chang}},
  \bibinfo{journal}{Phys. Rev. A} \textbf{\bibinfo{volume}{34}},
  \bibinfo{pages}{4550} (\bibinfo{year}{1986}).

\bibitem[{\citenamefont{Fischer et~al.}(1997)\citenamefont{Fischer, Godefroid,
  and Olsen}}]{key-Fischer1997}
\bibinfo{author}{\bibfnamefont{C.}~\bibnamefont{Fischer}},
  \bibinfo{author}{\bibfnamefont{M.}~\bibnamefont{Godefroid}},
  \bibnamefont{and} \bibinfo{author}{\bibfnamefont{J.}~\bibnamefont{Olsen}},
  \bibinfo{journal}{J. Phys. B} \textbf{\bibinfo{volume}{30}},
  \bibinfo{pages}{1163} (\bibinfo{year}{1997}).

\bibitem[{\citenamefont{Galvez et~al.}(2003)\citenamefont{Galvez, Buendia, and
  Sarsa}}]{key-Galvez2003}
\bibinfo{author}{\bibfnamefont{F.}~\bibnamefont{Galvez}},
  \bibinfo{author}{\bibfnamefont{E.}~\bibnamefont{Buendia}}, \bibnamefont{and}
  \bibinfo{author}{\bibfnamefont{A.}~\bibnamefont{Sarsa}}, \bibinfo{journal}{J.
  Chem. Phys.} \textbf{\bibinfo{volume}{118}}, \bibinfo{pages}{6858}
  (\bibinfo{year}{2003}).

\bibitem[{\citenamefont{Cheng et~al.}(1979)\citenamefont{Cheng, Kim, and
  Desclaux}}]{key-Cheng1979}
\bibinfo{author}{\bibfnamefont{K.}~\bibnamefont{Cheng}},
  \bibinfo{author}{\bibfnamefont{Y.}~\bibnamefont{Kim}}, \bibnamefont{and}
  \bibinfo{author}{\bibfnamefont{J.}~\bibnamefont{Desclaux}},
  \bibinfo{journal}{At. Data Nucl. Data Tables} \textbf{\bibinfo{volume}{24}},
  \bibinfo{pages}{111} (\bibinfo{year}{1979}).

\bibitem[{\citenamefont{Ynnerman and Froese-Fisher}(1995)}]{key-Ynnerman1995}
\bibinfo{author}{\bibfnamefont{A.}~\bibnamefont{Ynnerman}} \bibnamefont{and}
  \bibinfo{author}{\bibfnamefont{C.}~\bibnamefont{Froese-Fisher}},
  \bibinfo{journal}{Phys. Rev. A} \textbf{\bibinfo{volume}{51}},
  \bibinfo{pages}{2020} (\bibinfo{year}{1995}).

\bibitem[{\citenamefont{Jonsson and Fischer}(1997)}]{key-Jonsson1997}
\bibinfo{author}{\bibfnamefont{P.}~\bibnamefont{Jonsson}} \bibnamefont{and}
  \bibinfo{author}{\bibfnamefont{C.}~\bibnamefont{Fischer}},
  \bibinfo{journal}{J. Phys. B} \textbf{\bibinfo{volume}{30}},
  \bibinfo{pages}{5861} (\bibinfo{year}{1997}).

\bibitem[{\citenamefont{Lindroth and Hvarfner}(1992)}]{key-Lindroth1992}
\bibinfo{author}{\bibfnamefont{E.}~\bibnamefont{Lindroth}} \bibnamefont{and}
  \bibinfo{author}{\bibfnamefont{J.}~\bibnamefont{Hvarfner}},
  \bibinfo{journal}{Phys. Rev. A} \textbf{\bibinfo{volume}{45}},
  \bibinfo{pages}{2771} (\bibinfo{year}{1992}).

\bibitem[{\citenamefont{Chen et~al.}(2001)\citenamefont{Chen, Cheng, and
  Johnson}}]{key-Chen2001}
\bibinfo{author}{\bibfnamefont{M.}~\bibnamefont{Chen}},
  \bibinfo{author}{\bibfnamefont{K.}~\bibnamefont{Cheng}}, \bibnamefont{and}
  \bibinfo{author}{\bibfnamefont{W.}~\bibnamefont{Johnson}},
  \bibinfo{journal}{Phys. Rev. A} \textbf{\bibinfo{volume}{64}},
  \bibinfo{pages}{042507} (\bibinfo{year}{2001}).

\bibitem[{\citenamefont{Chen and Cheng}(1997)}]{key-Chen1997}
\bibinfo{author}{\bibfnamefont{M.}~\bibnamefont{Chen}} \bibnamefont{and}
  \bibinfo{author}{\bibfnamefont{K.}~\bibnamefont{Cheng}},
  \bibinfo{journal}{Phys. Rev. A} \textbf{\bibinfo{volume}{55}},
  \bibinfo{pages}{3440} (\bibinfo{year}{1997}).

\bibitem[{\citenamefont{Savukov and Johnson}(2002)}]{key-Savukov2002}
\bibinfo{author}{\bibfnamefont{I.}~\bibnamefont{Savukov}} \bibnamefont{and}
  \bibinfo{author}{\bibfnamefont{W.}~\bibnamefont{Johnson}},
  \bibinfo{journal}{Phys. Rev. A} \textbf{\bibinfo{volume}{65}},
  \bibinfo{pages}{042503} (\bibinfo{year}{2002}).

\bibitem[{\citenamefont{Safronova et~al.}(1996)\citenamefont{Safronova,
  Johnson, and Safronova}}]{key-Safronova1996}
\bibinfo{author}{\bibfnamefont{M.}~\bibnamefont{Safronova}},
  \bibinfo{author}{\bibfnamefont{W.}~\bibnamefont{Johnson}}, \bibnamefont{and}
  \bibinfo{author}{\bibfnamefont{U.}~\bibnamefont{Safronova}},
  \bibinfo{journal}{Phys. Rev. A} \textbf{\bibinfo{volume}{53}},
  \bibinfo{pages}{4036} (\bibinfo{year}{1996}).

\bibitem[{\citenamefont{Blundell et~al.}(1990)\citenamefont{Blundell, Johnson,
  and Sapirstein}}]{key-Blundell1990}
\bibinfo{author}{\bibfnamefont{S.}~\bibnamefont{Blundell}},
  \bibinfo{author}{\bibfnamefont{W.}~\bibnamefont{Johnson}}, \bibnamefont{and}
  \bibinfo{author}{\bibfnamefont{J.}~\bibnamefont{Sapirstein}},
  \bibinfo{journal}{Phys. Rev. A} \textbf{\bibinfo{volume}{42}},
  \bibinfo{pages}{3751} (\bibinfo{year}{1990}).

\bibitem[{\citenamefont{Sucher}(1980)}]{key-Sucher1980}
\bibinfo{author}{\bibfnamefont{J.}~\bibnamefont{Sucher}},
  \bibinfo{journal}{Phys. Rev. A} \textbf{\bibinfo{volume}{22}},
  \bibinfo{pages}{348} (\bibinfo{year}{1980}).

\bibitem[{\citenamefont{Mittleman}(1971)}]{key-Mittleman1971}
\bibinfo{author}{\bibfnamefont{M.}~\bibnamefont{Mittleman}},
  \bibinfo{journal}{Phys. Rev. A} \textbf{\bibinfo{volume}{4}},
  \bibinfo{pages}{893} (\bibinfo{year}{1971}).

\bibitem[{\citenamefont{Mittleman}(1972)}]{key-Mittleman1972}
\bibinfo{author}{\bibfnamefont{M.}~\bibnamefont{Mittleman}},
  \bibinfo{journal}{Phys. Rev. A} \textbf{\bibinfo{volume}{5}},
  \bibinfo{pages}{2395} (\bibinfo{year}{1972}).

\bibitem[{\citenamefont{Mittleman}(1981)}]{key-Mittleman1981}
\bibinfo{author}{\bibfnamefont{M.}~\bibnamefont{Mittleman}},
  \bibinfo{journal}{Phys. Rev. A} \textbf{\bibinfo{volume}{24}},
  \bibinfo{pages}{1167} (\bibinfo{year}{1981}).

\bibitem[{\citenamefont{Lindgren and
  Morrison}(1986)}]{key-Lindgren-Morrison1986}
\bibinfo{author}{\bibfnamefont{I.}~\bibnamefont{Lindgren}} \bibnamefont{and}
  \bibinfo{author}{\bibfnamefont{J.}~\bibnamefont{Morrison}},
  \emph{\bibinfo{title}{Atomic Many-Body Theory}}
  (\bibinfo{publisher}{Springer-Verlag}, \bibinfo{address}{Berlin},
  \bibinfo{year}{1986}), \bibinfo{edition}{2nd} ed.

\bibitem[{key(2006)}]{key-NIST2006}
\bibinfo{type}{Tech. Rep.}, \bibinfo{institution}{National Institute of
  Standards and Technology} (\bibinfo{year}{2006}), \bibinfo{note}{uRL:
  http://physics.nist.gov/PhysRefData/ASD/index.html}.

\bibitem[{\citenamefont{Cheng et~al.}(1993)\citenamefont{Cheng, Johnson, and
  Sapirstein}}]{key-Cheng1993}
\bibinfo{author}{\bibfnamefont{K.}~\bibnamefont{Cheng}},
  \bibinfo{author}{\bibfnamefont{W.}~\bibnamefont{Johnson}}, \bibnamefont{and}
  \bibinfo{author}{\bibfnamefont{J.}~\bibnamefont{Sapirstein}},
  \bibinfo{journal}{Phys. Rev. A} \textbf{\bibinfo{volume}{47}},
  \bibinfo{pages}{1817} (\bibinfo{year}{1993}).

\bibitem[{\citenamefont{Sapirstein}(1993)}]{key-Sapirstein1993}
\bibinfo{author}{\bibfnamefont{J.}~\bibnamefont{Sapirstein}},
  \bibinfo{journal}{Phys. Scr.} \textbf{\bibinfo{volume}{46}},
  \bibinfo{pages}{52} (\bibinfo{year}{1993}).

\bibitem[{\citenamefont{Chaudhuri et~al.}(1998)\citenamefont{Chaudhuri, Das,
  and Freed}}]{key-Chaudhuri1998}
\bibinfo{author}{\bibfnamefont{R.}~\bibnamefont{Chaudhuri}},
  \bibinfo{author}{\bibfnamefont{B.}~\bibnamefont{Das}}, \bibnamefont{and}
  \bibinfo{author}{\bibfnamefont{K.}~\bibnamefont{Freed}}, \bibinfo{journal}{J.
  Chem. Phys.} \textbf{\bibinfo{volume}{108}}, \bibinfo{pages}{2556}
  (\bibinfo{year}{1998}).

\end{thebibliography}

\end{document}